\documentclass[conference,compsoc]{IEEEtran}
\IEEEoverridecommandlockouts
\usepackage[nolist]{acronym}
\usepackage{tabularx}
\usepackage{booktabs}
\usepackage{todonotes}
\usepackage{caption}
\usepackage{subcaption}
\usepackage{multirow}
\usepackage{graphicx}
\begin{acronym}[AoI]
\acro{AoI}{Area of Interest}
\end{acronym}
\begin{acronym}[ARQ]
\acro{ARQ}{Automatic Repeat Request}
\end{acronym}
\begin{acronym}[CACC]
\acro{CACC}{Cooperative Adaptive Cruise Control}
\end{acronym}
\begin{acronym}[CAM]
\acro{CAM}{Cooperative Awareness Message}
\end{acronym}
\begin{acronym}[CBR]
\acro{CBR}{Channel Busy Ratio}
\end{acronym}
\begin{acronym}[CA]
\acro{CA}{Cooperative Awareness}
\end{acronym}
\begin{acronym}[CBF]
\acro{CBF}{Contention-Based Forwarding}
\end{acronym}
\begin{acronym}[CCAM]
\acro{CCAM}{Cooperative, Connected and Automated Mobility}
\end{acronym}
\begin{acronym}[C-ITS]
\acro{C-ITS}{Cooperative Intelligent Transport Systems}
\end{acronym}
\begin{acronym}[CP]
\acro{CP}{Collective Perception}
\end{acronym}
\begin{acronym}[CPM]
\acro{CPM}{Collective Perception Message}
\end{acronym}
\begin{acronym}[DCC]
\acro{DCC}{Decentralized Congestion Control}
\end{acronym}
\begin{acronym}[DEN]
\acro{DEN}{Decentralized Environmental Notification}
\end{acronym}
\begin{acronym}[DENM]
\acro{DENM}{Decentralized Environmental Notification Message}
\end{acronym}
\begin{acronym}[DPD]
\acro{DPD}{Duplicate Packet Detection}
\end{acronym}
\begin{acronym}[DPL]
\acro{DPL}{Duplicate Packet List}
\end{acronym}
\begin{acronym}[ETSI]
\acro{ETSI}{European Telecommunications Standards Institute}
\end{acronym}
\begin{acronym}[FCD]
\acro{FCD}{Floating Car Data}
\end{acronym}
\begin{acronym}[ITS]
\acro{ITS}{Intelligent Transport Systems}
\end{acronym}
\begin{acronym}[LocT]
\acro{LocT}{Location Table}
\end{acronym}
\begin{acronym}[LocTE]
\acro{LocTE}{Location Table entry}
\end{acronym}
\begin{acronym}[MC]
\acro{MC}{Maneuver Coordination}
\end{acronym}
\begin{acronym}[MCM]
\acro{MCM}{Maneuver Coordination Message}
\end{acronym}
\begin{acronym}[mmWave]
\acro{mmWave}{millimiter wave}
\end{acronym}
\begin{acronym}[PAM]
\acro{PAM}{Platooning Awareness Message}
\end{acronym}
\begin{acronym}[PCM]
\acro{PCM}{Platooning Control Message}
\end{acronym}
\begin{acronym}[PDR]
\acro{PDR}{Packet-delivery Ratio}
\end{acronym}
\begin{acronym}[RadCom]
\acro{RadCom}{Radar-Based Communication}
\end{acronym}
\begin{acronym}[RSU]
\acro{RSU}{Road-side Unit}
\end{acronym}
\begin{acronym}[RHW]
\acro{RHW}{Road Hazard Warning}
\end{acronym}
\begin{acronym}[SHB]
\acro{SHB}{Single-Hop Broadcasting}
\end{acronym}
\begin{acronym}[S-CBR]
\acro{S-CBR}{Smoothed Channel Busy Ratio}
\end{acronym}
\begin{acronym}[TC]
\acro{TC}{Traffic Class}
\end{acronym}
\begin{acronym}[TDC]
\acro{TDC}{Transmit Data Rate Control}
\end{acronym}
\begin{acronym}[TPC]
\acro{TPC}{Transmit Power Control}
\end{acronym}
\begin{acronym}[TRC]
\acro{TRC}{Transmit Rate Control}
\end{acronym}
\begin{acronym}[TTL]
\acro{TTL}{Time-to-Live}
\end{acronym}
\begin{acronym}[VANET]
\acro{VANET}{Vehicular ad hoc Network}
\end{acronym}
\begin{acronym}[V2X]
\acro{V2X}{Vehicle-to-anything}
\end{acronym}
\begin{acronym}[V2V]
\acro{V2V}{Vehicle-to-Vehicle}
\end{acronym}
\ifCLASSOPTIONcompsoc
  \usepackage[nocompress,noadjust]{cite}
\else
  \usepackage{cite}
\fi
\ifCLASSINFOpdf
\else
\fi
\begin{document}

% Alternative title "Improving the safety and efficiency in 5.9GHz V2X platooning applications by offloading to Radar Communication"
\title{Offloading platooning applications from 5.9\,GHz V2X to Radar Communications: effects on safety and efficiency.\thanks{This work was partially supported by SAFER in the project ``Human Factors, Risks and Optimal Performance in Cooperative, Connected and Automated Mobility'', the Knowledge Foundation in the project “SafeSmart – Safety of Connected Intelligent Vehicles in Smart Cities” (2019-2024), and the ELLIIT Strategic Research Network in the project ``6G wireless'' – sub-project ``vehicular communications'', and Vinnova grant 2021-02568}}

\author{\IEEEauthorblockN{Elena Haller, Galina Sidorenko, Oscar Amador, Emil Nilsson}
\IEEEauthorblockA{School of Information Technology\\
Halmstad University\\
Halmstad, Sweden 30118\\
Emails: \{elena.haller,galina.sidorenko,oscar.molina,emil.nilsson\}@hh.se}
%\and
%\IEEEauthorblockN{Galina Sidorenko}
%\IEEEauthorblockA{School of Information\\Technology\\
%Halmstad University\\
%Halmstad, Sweden 30118\\
%Email: bb@hh.se}
%\and
%\IEEEauthorblockN{Oscar Amador}
%\IEEEauthorblockA{School of Information\\Technology\\
%Halmstad University\\
%Halmstad, Sweden 30118\\
%Email: cc@hh.se}
%\and
%\IEEEauthorblockN{Emil Nilsson}
%\IEEEauthorblockA{School of Information\\Technology\\
%Halmstad University\\
%Halmstad, Sweden 30118\\
%Email: dd@hh.se}
}


% compsoc conferences compared to traditional conferences), use this
%Telephone: (800) 555--1212, Fax: (888) 555--1212}


\maketitle

\begin{abstract}
V2X communications are nowadays performed at 5.9\,GHz spectrum, either using WiFi-based or Cellular technology. The channel capacity is limited, and congestion control regulates the number of messages that can enter the medium. With user rate growing, overloading becomes a factor that might affect road safety and traffic efficiency. The present paper evaluates the potential of using Radar-Based Communication (RadCom) for offloading the V2X spectrum. We consider a heavy-duty vehicle (HDV) platooning scenario as a case of maneuver coordination where local messages are transmitted by means of RadCom at different penetration rates. Simulations show significant improvements in channel occupation and network reliability. As a result, RadCom allows for shorter safe and energy efficient inter-vehicle distances.  \end{abstract}


% \fi
\IEEEpeerreviewmaketitle

\section{Introduction}
\IEEEPARstart{M}{obility} with maximized road safety and traffic efficiency is a goal reflected in the United Nation's Social Development Goals~\cite{Agenda2030}, and in global initiatives such as Vision Zero~\cite{ecVisionZero}, adopted by the European Commission. The road towards future mobility in Vision Zero is divided in stages called \textit{Days}, which go from Day 1 (where we currently are) until Day 4, when \ac{CCAM} is expected to be present on all roads and at all times.

Efforts to achieve full the final stage of \ac{CCAM} occur in two main fronts: automated mobility and \ac{C-ITS}. The latter uses vehicular communications, such as \acp{VANET}, that use access technologies such as 802.11p. For Day 1 services, aimed at increasing the awareness of road users, there are technologies such as the \ac{CA}~\cite{etsiCAM} and \ac{DEN}~\cite{etsiDENM} basic services, which are used in the framework defined by the \ac{ETSI}. These services rely on the exchange of messages that inform road users about each other's presence (through \acp{CAM}) or about risks on the road (using \acp{DENM}). These messages are exchanged using access technologies such as ETSI~ITS-G5, which is based on 802.11p.

Further Days are also expected to rely on messages. Day 2, when \textit{cooperation} starts, uses \acp{CPM} to exchange information about detected objects~\cite{Delooz:2020}. On Day 3, road users share their intentions (e.g., desired trajectories), and finally, on Day 4, vehicles coordinate their maneuvers. These features are expected to be performed by the \ac{MC} service, powered by \acp{MCM}.

Early forms of maneuver coordination are \ac{CACC} and platooning. These applications are also based on messages. For example, platooning uses \acp{CAM} and \acp{PAM} to inform neighbors of the ability to form a platoon and negotiate its start (sent in broadcast mode), and \acp{PCM} to maintain the platoon (sent in unicast between platoon members)~\cite{etsiMCO}.

This means that, from Day 1, the medium will be occupied by a myriad of messages to power services. Even for Day 1, the existence of traffic with different characteristics and priorities can cause problems in efficiency and effectiveness for safety applications (\cite{OscarGOT,Delooz:2020}). These issues stem from the capacity of the channel to accommodate a certain number of messages before reaching congestion (\cite{BaiocchiAoI,AmadorAccessDCC}). Thus, if more messages enter the system, e.g., \acp{MCM}, \acp{PCM}, the ability of the channel to accommodate them will be hindered.

This paper proposes the use of \ac{RadCom} as an alternative channel to offload future mobility use cases, such as platooning. We present a study of the potential saving in terms of medium usage (\ac{CBR}), and whether such offloading allows reaching the goal of platooning and \acp{C-ITS} in general: minimizing risks of accidents (e.g., collisions between road users) and increasing efficiency. 

The contributions of this work are:
\begin{enumerate}
    \item A simulation-based study of the effect of offloading \ac{V2V} communications from the \ac{C-ITS} medium to bumper-to-bumper \ac{RadCom}.
    \item An analysis of the effect of this offloading on minimizing safe inter-vehicle distances in a platoon.
    \item An analysis of the effect of these distances on traffic and in-vehicle efficiency. 
\end{enumerate}

The rest of the paper is organized as follows: in Section~\ref{sec:related_work}, we explore the related work on platooning and its performance using \acp{VANET}; Section~\ref{sec:proposal} presents our proposal for \ac{RadCom}-enabled channel offloading; Section~\ref{sec:experiment} presents an experimental evaluation of the system; Section~\ref{sec:discussion} presents an analysis of our results and its effects on road safety and traffic efficiency; and finally Section~\ref{sec:conclusion} presents the conclusions and our future lines of work.

\section{Platooning}
\label{sec:related_work}

%\subsection{Platooning}

Platooning has been widely studied in the context of safety and efficiency. An early example of electronics-assisted platooning is the "Electronic Tow Bar" resulting from the European project PROMOTE-CHAUFFEUR~\cite{bonnet2000fuel}. Further efforts from the industry and research communities are reflected in brand-specific projects SARTRE~\cite{sartre} and COMPANION~\cite{companion}, as well as multi-brand projects like ENSEMBLE~\cite{ensemble}. In this section, we describe network-enabled platooning, as expected by ETSI and as studied by the research community.

\subsection{Network-enabled platooning}
Platooning is an \ac{C-ITS} application contemplated in the ETSI ITS framework~\cite{etsiMCO}. As it is for Day 1 applications, it relies on the exchange of messages: \acp{CAM} inform about a vehicle's capability to platoon, \acp{PAM} are used to negotiate platoon creation, and \acp{PCM} are sent between the leader and members (in unicast) to maintain the platoon. 

The platoon leader exchanges information with the rest of the members and with other road users. This enables functionalities for platoon safety and efficiency, such as negotiating inter-vehicle distances to enable safe braking in emergency situations~\cite{GalinaDist}, or exchanging information on attributes and capabilities (e.g., acceleration/deceleration capability).

While the platoon is enabled, the rest of \ac{V2X} communications are still used, so there is a possibility that \acp{PCM} use a channel different from the main \ac{V2X} channel (e.g., CCH for ETSI ITS-G5), which is used by other safety-critical applications. However, even if a different channel is used, they are likely to suffer from access layer phenomena, as we explore in Section~\ref{subsub:access_phenomena}.

\subsection{Access phenomena affecting platooning}
\label{subsub:access_phenomena}
\begin{figure}
    \centering
    \includegraphics[width=\linewidth]{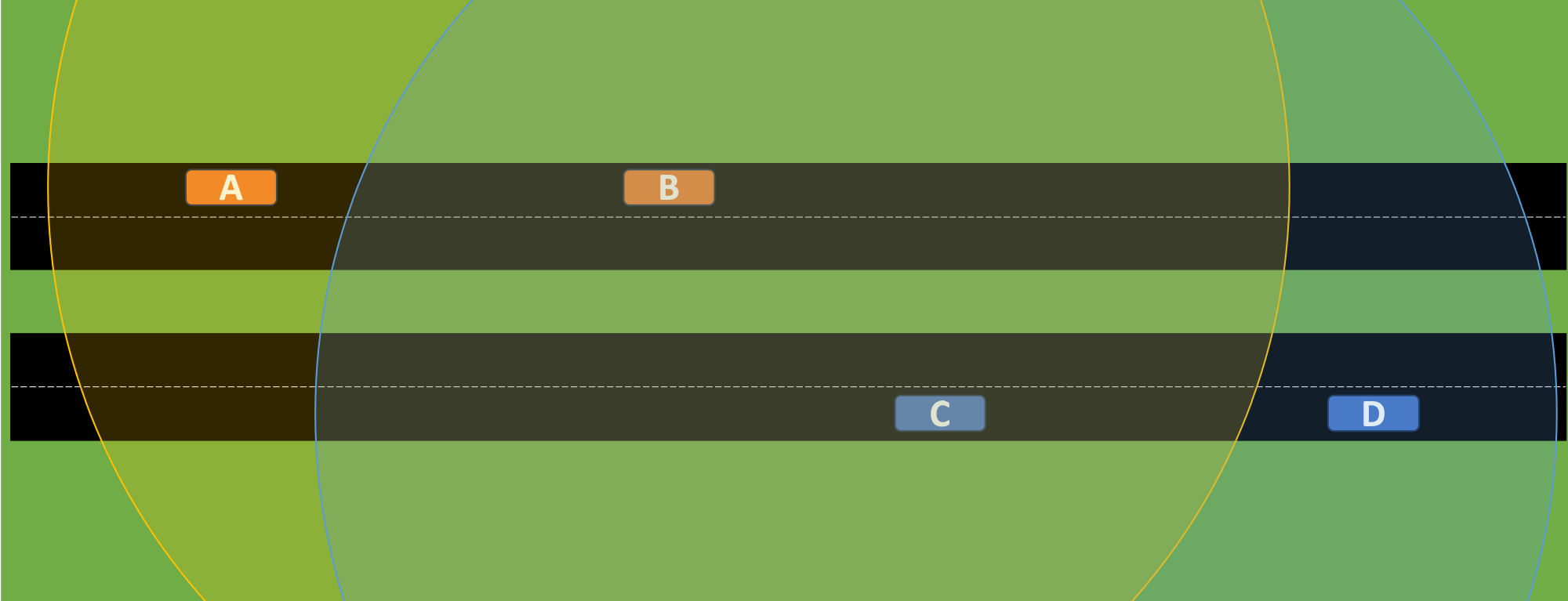}
    \caption{Hidden and exposed node phenomena in V2X scenarios}
    \label{fig:hidden-exposed}
\end{figure}

There are several Access layer phenomena that affect V2X-enabled platooning. Some of these phenomena are inherent to the nature of WiFi-based protocols such as ETSI~ITS-G5, such as hidden and exposed nodes; and others are related to channel congestion and how different frameworks deal with it. Fig.~\ref{fig:hidden-exposed} summarizes the hidden and exposed node phenomena. A and D are hidden nodes for C and B, respectively, and messages they send to each other (A and C) could collide with messages coming from the hidden nodes. Similarly, if A and B want to communicate with each other, communication between C and D can prevent them from accessing the medium. This is the exposed node phenomenon. While there are mechanisms, e.g., back-off procedures in WiFi networks, to counter hidden and exposed nodes, there are access protocols in C-ITS that only use them when unicast communication takes place (e.g., ETSI ITS-G5 does not use exponential back-off for single-hop broadcast traffic, but does so for unicast)~\cite{etsiCoex}.

An analysis of \ac{CAM}-based platooning is performed in~\cite{NikitaCamPlatooning}. Here, authors present an evaluation of the \ac{CA} basic service as an enabler for platooning and identify a problem with message synchronization. While \ac{CAM} synchronization is solved with kinematic generation, \acp{PCM} are generated at high, periodic rates~\cite{etsiMCO}, and are thus susceptible to collisions due to synchronization.

Furthermore, even if \acp{PCM} are sent in unicast mode (as opposed to \acp{CAM} and \acp{DENM}, which are broadcast), neighbors can overhear these exchanges (i.e., sense the medium as occupied). In moderate to high density scenarios, these exchanges add to the existing channel occupation and create congestion. The work in~\cite{delftmultilane} analyzes the performance of ETSI DCC in multi-lane platooning scenarios, and proposes the use of congestion control techniques different to adapting message rates (e.g., controlling transmission power).

There is a need for "intra-platoon" communications to occur without interfering with other applications sharing the medium. The use of \ac{mmWave} communications for platooning applications is explored in~\cite{mmWavePlatoon}. The authors use \ac{mmWave} to rely sensor information using multi-hop dissemination. The difference with our proposal is that, while they consider a generic \ac{mmWave} antenna, we analyze the possibility of piggy-backing communications specifically on radars with specific attributes and capabilities.

\section{RadCom-enabled Platooning}
\label{sec:proposal}

\begin{figure}[tb]
    \centering
    \begin{subfigure}[b]{\linewidth}
      \centering
      \caption{Network-enabled platooning. C-ITS spectrum}
      \includegraphics[width=1\textwidth]{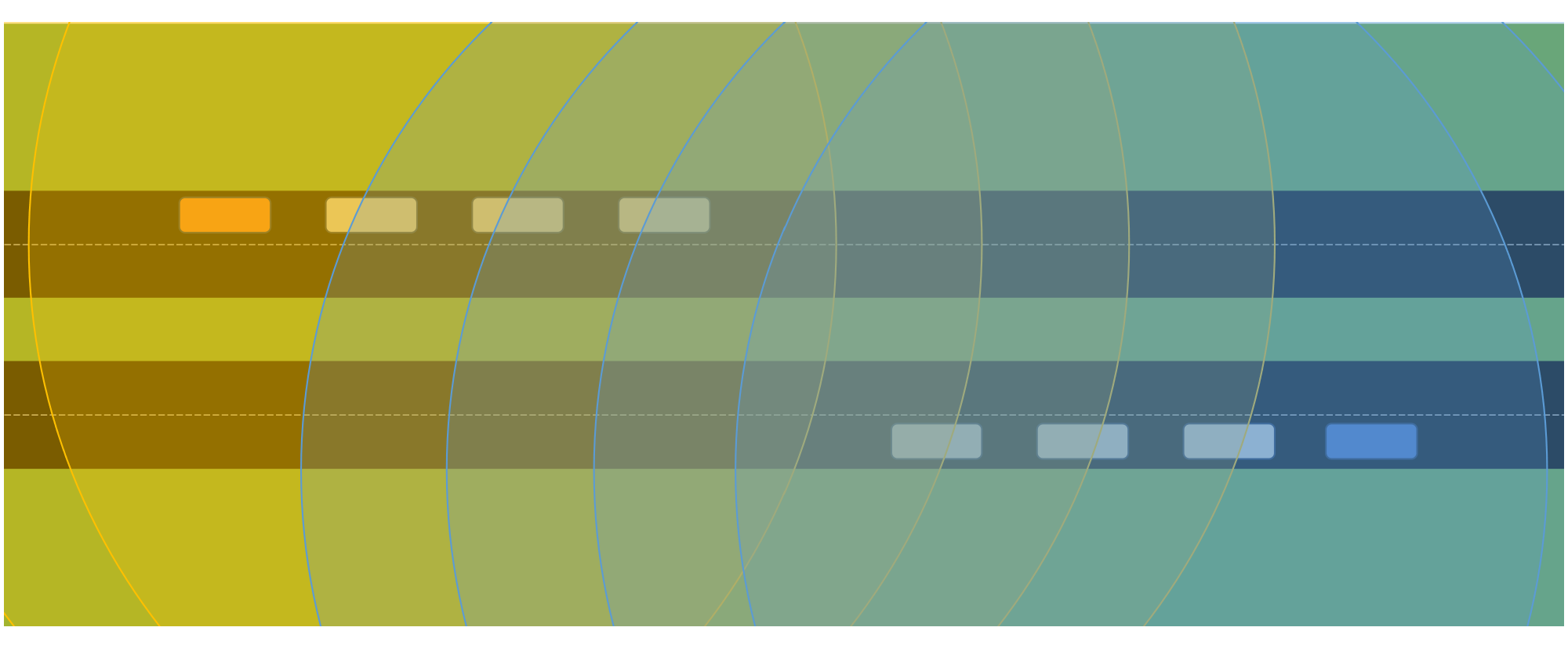}
      \label{fig:broadcast}
    \end{subfigure}
     \begin{subfigure}[b]{\linewidth}
      \centering
      \caption{Network-enabled platooning. C-ITS spectrum and RadCom}
      \includegraphics[width=1\textwidth]{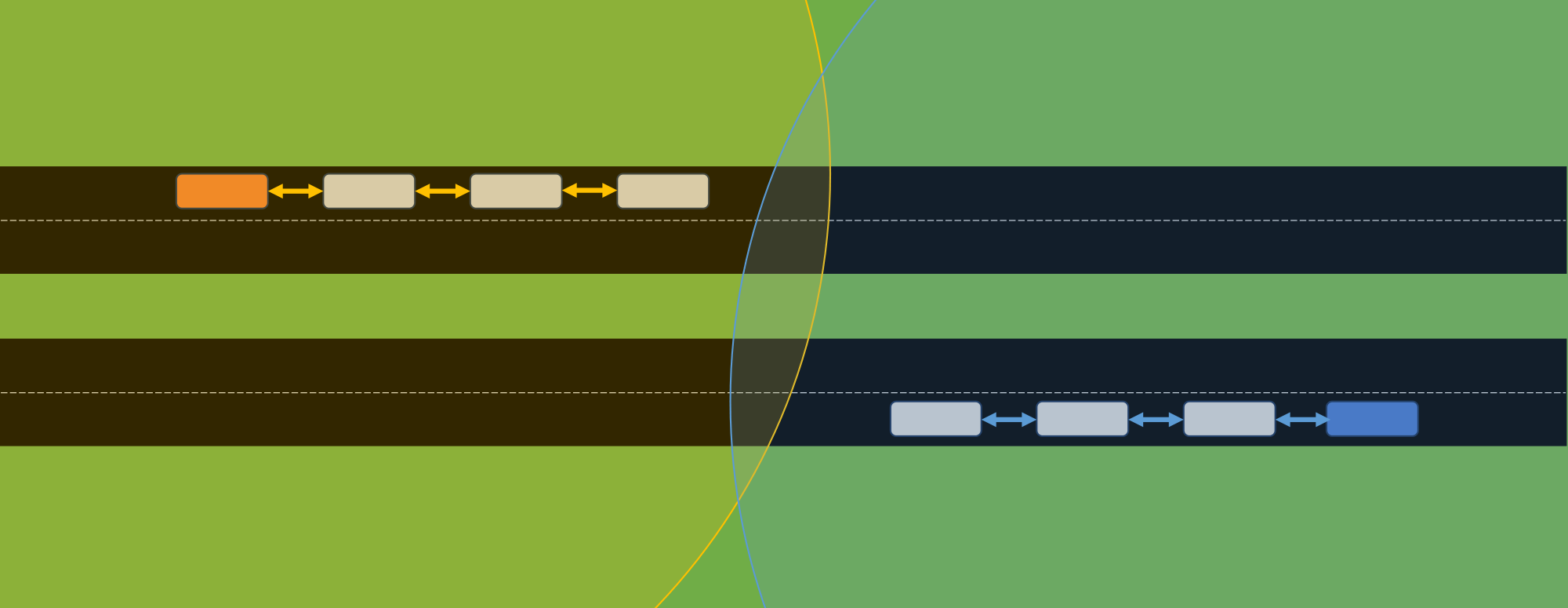}
      \label{fig:hybrid}
    \end{subfigure}
    \caption{Representation of offloading intra-platoon communications to RadCom. The yellow and blue shades represent the coverage area for the members of each platoon, yellow for the westbound and blue for the eastbound.}
    \label{fig:platoons_ranges}
\end{figure}

Fig.~\ref{fig:platoons_ranges} summarizes our proposal to offload intra-platoon communications to \ac{RadCom}. The top part of the figure (\ref{fig:broadcast}) shows the foreseeable status of network-enabled platooning. After negotiating the start of a platoon using \acp{CAM} and \acp{PAM}, the platoon leaders (dark nodes) start communicating with the members (light nodes) exchaning \acp{PCM} in unicast mode. The arches express a conservative range for the wireless signals from each node (yellow for the platoon on the left and blue for the one on the right). These ranges are typically measured above 300\,m, but are sometimes greater~\cite{Paulin2015}.

This means that, in scenarios like this, intra-platoon communications will not only interfere with other platoon members, but also with other platoons and even other vehicular communications. Fig.~\ref{fig:broadcast} shows that the tails of both platoons are within the range of most of the nodes, and are likely to suffer from the exposed node phenomenon, as described in Section~\ref{subsub:access_phenomena}.

Fig.~\ref{fig:hybrid} shows how intra-platoon communications can be offloaded to \ac{V2V} \ac{RadCom}. On this example, the leaders can send commands, e.g., through \acp{PCM}, and receive feedback or other communications from platoon members through multi-hop \ac{RadCom} (represented by arrows). Then, radio resources would be available for other applications, e.g., inter-platoon communications. At the very least, this offloading will avoid causing congestion in the \ac{C-ITS} spectrum.

\subsection{RadCom ability to support V2V communications}

The bare minimum requirements for \ac{RadCom} to support \ac{V2V} communications are the specifications for the ETSI ITS-G5 medium~\cite{etsiG5specs}. These are:

\begin{enumerate}
    \item \textbf{Data rate:} support 3\,Mbit/s, 6\,Mbit/s and 12\,Mbit/s. The default rate in ETSI ITS-G5 is 6\,Mbit/s.
    \item \textbf{Message rate:} maximum 40\,Hz and minimum 1\,Hz.
\end{enumerate}

The body of work on \ac{mmWave} systems supporting vehicular applications shows that these requirements can be met by generic \ac{mmWave} deployments. The work in~\cite{mmWavePlatoon} tests their proposed multi-hop system at Gbit/s rates when exchanging sensor information. The work in~\cite{mmWaveCapacity} assesses the capacity of generic \ac{mmWave} when coding errors are present. OFDM waveforms has been proposed for automotive radar enabling wide bandwidth communication~\cite{9266449}. Communication networks may also be instrumental in avoiding interference between wide band radar sensors~\cite{9266520,9455180,9127843}. Their results show that, starting with bandwidths of a fraction of a GHz, rates in tens of Mbit/s are possible in line-of-sight scenarios. Thus, we can expect \ac{RadCom} to be able to accommodate the requirements that are set for ETSI ITS-G5 and beyond.

\section{Evaluation of Channel Occupation}
\label{sec:experiment}

We use Artery~\cite{Artery} as our simulation toolkit. It combines OMNET++ and Vanetza --- a C++ implementation of the ETSI ITS protocol stack. Our setup uses Artery's integration with Veins~\cite{Veins} for the physical layer. Finally, SUMO~\cite{sumo2012} provides the mobility model for the road topology. Simulation parameters are specified in Table~\ref{tbl:simpars}.

\begin{table}[tbh!]
	\centering
	\caption{Simulation Parameters}
	\label{tbl:simpars}
	\begin{tabular}{| l | l |}
		\hline
		\textbf{Parameter}  & \textbf{Values} \\
		\hline
		Access Layer protocol & ITS-G5 (IEEE 802.11p) \\
		Channel bandwidth & 10\,MHz at 5.9\,GHz \\
		Data rate & 6\,Mbit/s \\
		Transmit power & 20\,mW \\
		Path loss model & Simple Path-loss Model \\
		Maximum transmission range & 1500\,m \\
		CAM packet size & 285 bytes \\
		CAM Traffic Class & TC2 \\
            CAM generation frequency & Kinematic-based~\cite{etsiCAM} \\
		PCM packet size & 301 bytes \\
		PCM Traffic Class & TC1 \\
		PCM generation frequency & Periodic at 2\,Hz\\
		RadCom penetration rate & 0, 50, 100\%\\
		\hline
	\end{tabular}
\end{table}

\subsection{Simulation Scenario}
\label{subsec:scenario}

For our scenario, we simulate a 5\,km long road with four lanes in each direction. Vehicles occupy the road with a density of 30\,veh/km per lane and are running in a steady state, where we can consider them to have organized platoons of different lengths, and have different roles (platoon leader and platoon member). We take measurements for 30\,s after a warm-up period of 120\,s. Vehicles send \acp{CAM} (generated dynamically following ETSI rules~\cite{etsiCAM}), and \acp{PCM} that work for controlling and maintaining the platoon (generated periodically every 500\,ms). We send both messages on the same ETSI ITS-G5 channel, with \acp{PCM} having higher priority than \ac{CAM} since we consider them to be more critical, although there is not a standardized priority for \acp{PCM} as of now~\cite{etsiMCO}.

A subset of platoons offload \acp{PCM} to bumper-to-bumper \ac{RadCom}. This means that these vehicles stop sending \acp{PCM} on the ETSI ITS-G5 spectrum and perform platoon control and maintenance "bumper-to-bumper". We increase the number of \ac{RadCom} platoons (i.e., penetration rate) until only the platoon leaders send messages on the ETSI ITS-G5 channel (i.e., 100\% penetration rate).

We measure conditions in the ETSI ITS-G5 channel. The performance metrics we evaluate from the simulation are:
\begin{itemize}
    \item \textbf{\ac{PDR}}: the number of successful message receptions divided by the total expected receptions.
    \item \textbf{\ac{S-CBR}}: the smoothed average of \ac{CBR} measurements which is used in the \ac{DCC} mechanism~\cite{etsiNewDcc}.  
\end{itemize}

The results above allow the calculation of safety and efficiency metrics presented in Section~\ref{sec:fuel}. We use the work in~\cite{GalinaDist} to relate network performance to inter-vehicle distances, and then extrapolate them to fuel efficiency.

\subsection{Simulation results}
\label{subsec:simulation_results}

\begin{table}[tbh!]
	\centering
	\caption{Results for messages in the C-ITS spectrum at $d\leq200$m with different penetration rates of RadCom}
	\label{tbl:simres}
	\begin{tabular}{ c  l  l  r}
		\toprule
        \textbf{RadCom Penetration Rate}   & \textbf{PDR} & \textbf{S-CBR} & \textbf{Latency}\\
		\midrule
        0\% & 0.6985 & 0.6176 & 136.80~ms \\
        50\% & 0.7859 & 0.6119 & 109.57~ms \\
        100\% & 0.9015 & 0.2217 & 1.45~ms \\
		\bottomrule
	\end{tabular}
\end{table}

Table~\ref{tbl:simres} shows the results of our experiment. The table shows the average \ac{PDR} and latency for \acp{PCM}. For the 100\% rate, only the leaders send messages on the \ac{C-ITS} spectrum and the rest of the platoon communicates using RadCom. Thus, the \ac{PDR} for \acp{PCM} is mostly affected by propagation phenomena and the interference from \acp{CAM}. Therefore, even sending command-and-control messages at a high rate, offloading to \ac{RadCom} allows platoon members to listen to leaders with significantly increased reliability.

It is worth noting that, even with half of the fleet not sending \acp{PCM}, channel occupation stays at similar levels as when all vehicles send \acp{PCM}. This number, slightly above 0.61 is close to the theoretical CBR limit for ETSI~ITS-G5 (0.68) and to the point where medium occupation converges ($CBR=0.65$)~\cite{AmadorAccessDCC}. While occupation stays similar, the effect of offloading to \ac{RadCom} is noticeable in PDR and in latency: more packets arrive, and they do so faster. Nevertheless, even at this vehicle density (30~veh/km per lane), the stress on the medium is noticeable. However, when only platoon leaders send \acp{PCM} (100\% penetration rate), besides the increased \ac{PDR}, the value for latency significantly better. Delays are minimal, and thus, more messages arrive, and they do so in a timely fashion.

\section{Optimal Distances and Fuel Consumption}
\label{sec:fuel}

\begin{figure}[ht!]
   \centering	
   \includegraphics[width=\linewidth]{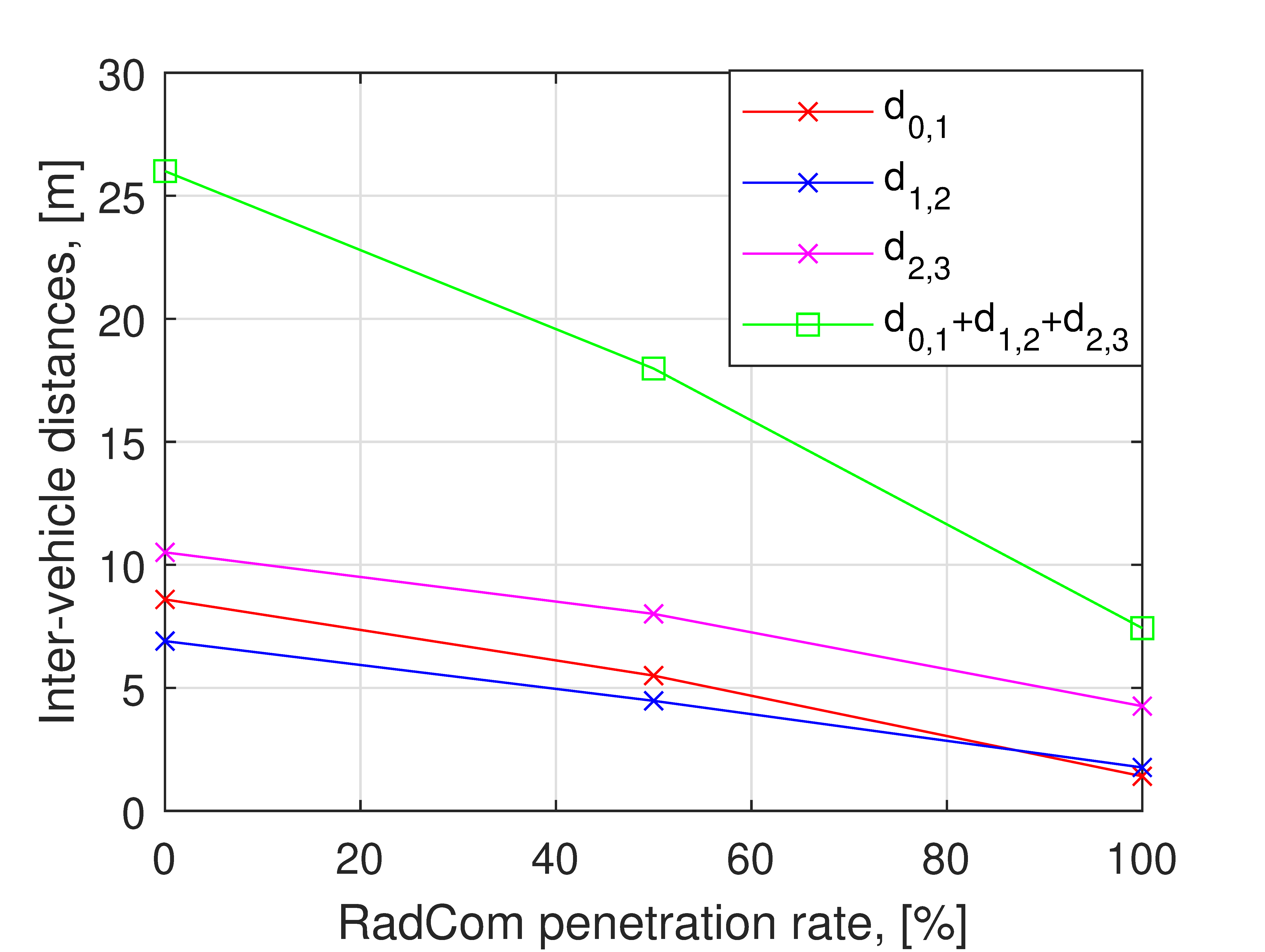}
  \caption{Safe inter-vehicle distances in a platoon of four vehicles versus RadCom penetration rate. Here, $d_{i-1,i}$ denotes an inter-vehicle distance between vehicles $i-1$ and $i$ where vehicle $0$ is the leader.}
   \vspace{0.2cm}
   \label{fig:dist_1}
\end{figure}

Increased offloading of V2X channel results in increased PDR (Table~\ref{tbl:simres}). Consequently, inter-vehicle distances within the platoon can be decreased without compromising safety. Fig.~\ref{fig:dist_1} shows how distances between vehicles of one platoon can be shortened with the increased RadCom penetration rate. The minimum safe inter-vehicle distances were calculated according to~\cite{GalinaDist} for a simulated scenario involving a four-vehicle platoon on a flat and straight road section. Reduced distances between vehicles contribute to a decrease in air drag force, subsequently resulting in reduced fuel consumption. In our simulation, the platoon of four heavy-duty trucks can save 2\% of fuel with 50\% RadCom penetration rate compared to the scenario when radar communications are not utilized. Furthermore, a 100\% offload of the V2X channel leads to even greater fuel efficiency, with a substantial 5.6\% reduction in fuel used. Lower fuel consumption leads to cost savings for individuals and businesses, as well as diminishes environmental impact.

% \section{Optimal distances and fuel consumption (Elena)}
% \label{sec:discussion}
% \begin{figure}[ht!]
%    \centering	\includegraphics[width=0.8\linewidth]{ex1_75.eps}
%   \caption{Inter-vehicle distances in a platoon of 4 vehicles versus RadCom penetration rate. }
%    \vspace{0.2cm}
%    \label{fig:example_2}
% \end{figure}

\section{Discussion}
\label{sec:discussion}
Below we discuss various issues related to using RadCom for intra-platooning communications. All of them represent possible  directions for further research.

%see if this can summarize the section
%, e.g., the effect on platooning for different types of HDVs and different road scenarios, protocol optimizations, and challenges for RadCom platooning.

The energy savings in Section \ref{sec:fuel} are presented for one particular scenario only. 
%and in reality are more complicated to compute.
The full mechanical model (\cite{KTH?}, Eq. (1,2)) takes into account road and vehicles' geometry (possible bumps, inclinations, curvatures) as well as their dynamical properties (variable friction and resistance coefficients). One might also consider possible thermal effects. Thus, for internal combustion engine (ICE) vehicles, too small distances imply additional heating that, in turn, increases fuel consumption \cite{THATreport}. However, this is not the case for electrical vehicles. 

Another factor that ambiguously affects energy costs is PCM rate. 
Low PCM rates lead to decreasing service costs (e.g., resource usage, and data transmission costs), whereas the high ones allow for shorter inter-vehicle distances. Thus, energy consumption is determined by the balance between low/high maintenance costs and savings related to shorter/longer distances.

%The rate of PCM by RadCom affects performance of a platoon in several ways.

%- Min distances heating vs fuel type

%Variable rate PCM
One of the points to evaluate is the pertinence of having \acp{PCM} sent at a fixed interval, as considered in~\cite{etsiMCO}. While these \textit{heartbeat} messages keep platoon members informed about the status of the cluster, there are situations where \ac{PCM} frequencies can be lowered (e.g., in flat, straight stretches of a highway).
%, while other situations might require them (e.g., starting a slope, going into a pronounced curve, changing road or weather conditions).
Further work is required in order 
to determine if variable \ac{PCM} rates are energy beneficial and whether RadCom can support such scenarios.

%in order to determine whether a variable \ac{PCM} rates are more suitable to accommodate the wide range of situations that can occur during a trip, and whether RadCom can support these scenarios.

The weather causes interference for radar in general and RadCom specifically. The detection capabilities of radars are affected by adverse weather conditions~\cite{adversephotonicradar, 9564436} which are likely to affect RadCom links as they do with certain cases of V2X communications~\cite{mmWaveAdverse}. Further exploration is needed to understand the impact of weather-related failures, assess potential mitigation strategies (e.g., adding re-transmission protocols with or without forward error correction), and evaluate their effects on throughput.

Another issue that can affect performance in general for RadCom-enabled platooning is security. The ETSI ITS framework specifies a security architecture~\cite{etsiSecArchitecture} with different requirements for each ITS service. For example, confidentiality and privacy requirements are different for \acp{CAM} and for \acp{DENM}, given their different nature. In \ac{DENM} cases, such as road hazard warnings, the trade-off between confidentiality and road safety is leveraged differently. Work has to be performed to assess the need to encrypt RadCom-exchanged \acp{PCM} since they have a different dissemination scheme than those exchanged using the C-ITS spectrum.

Finally, one of the contributions of this paper was the evaluation of the effect of RadCom penetration rates. The nature of the vehicular industry creates a phenomenon where the average age of a vehicle is 12 years for passenger vehicles and 14.2 years for trucks~\cite{fleetage}. This means that, even if 100\% of vehicles produced from today include RadCom nodes, it is unlikely that the penetration rate will reach 100\% before several decades pass. However, one solution could be to retrofit radar-equipped vehicles (especially heavy-duty vehicles) with nodes adapted to their currently existing radars. Further work on the effect of a mixed-capability fleet shall be performed.

\section{Conclusion and Future Work}
\label{sec:conclusion}

We presented a proposal to offload intra-platoon communications, which are expected to use \acp{PCM} sent in the 5.9\,GHz V2X spectrum, to RadCom. We measured the potential benefit of the use of this additional access technology in a less congested V2X spectrum when adoption rates are high. Thus, lower congestion is reflected in increased network reliability and reduced end-to-end delays for safety-critical messages.

This increased reliability allows reducing the minimum inter-vehicle distance required for safe platooning. Therefore, other efficiency metrics are also boosted as a consequence of offloading communications to RadCom. We showed that fuel efficiency is increased with reduced distances, and with this efficiency, it can be argued that other beneficial societal and economical impacts occur.

Finally, we elaborated on other issues affecting platooning in general and RadCom-enabled platooning specifically:
\begin{itemize}
    \item lowered distances effects on platoons of ICE and electric vehicles,
    \item fixed and dynamic \acp{PCM} rates,
    \item effects of adverse conditions on RadCom, and
    \item security requirements.
\end{itemize}
Future work includes the thorough evaluation of RadCom to comply with the requirements set for existing services and \textit{Future Mobility} services based on V2V communications, such as intention sharing and maneuver coordination.

%\ifCLASSOPTIONcompsoc
%\else
%\fi




%\bibliographystyle{IEEEtran}
\bibliographystyle{IEEEtran}
\bibliography{radcom}

\end{document}